\documentclass[prl,superscriptaddress,twocolumn,amsmath,amssymb,floatfix]{revtex4-1}
\usepackage{graphicx}

\begin{document}
\title{ 2D transport and screening in topological insulator surface states} 

\author{S. Adam}
\affiliation{Center for Nanoscale Science and Technology, National Institute of Standards and Technology, Gaithersburg, MD 20899, USA}
\author{E. H. Hwang}
\affiliation{Condensed Matter Theory Center, Department of Physics, University of Maryland, College Park, MD 20742-4111}
\author{S. Das Sarma}
\affiliation{Condensed Matter Theory Center, Department of Physics, University of Maryland, College Park, MD 20742-4111}

\date{\today}

%%%%%%%%%%%%%%%%% END OF PREAMBLE %%%%%%%%%%%%%%%%

\begin{abstract}  
We study disorder effects on the surface states of the topological
insulator Bi$_2$Se$_3$ close to the topologically
protected crossing point.  Close to charge neutrality, local
fluctuations in carrier density arising from the random charged
disorder in the environment result in electron and hole puddles
that dominate the electronic properties of these materials.  By
calculating the polarizability of the surface state using the random
phase approximation, and determining the characteristics of puddles
using the self-consistent approximation, we find that band asymmetry plays
a crucial role in determining experimentally measured quantities
including the conductivity and the puddle autocorrelation length.
\end{abstract}

\maketitle
\normalsize

% \paragraph*{Introduction}

Topological insulators (TIs) are a new class of materials
\cite{kn:hasan2010,kn:moore2010,kn:qi2011,kn:culcer2011} that are typically
distinguished by their robust metallic surface state encapsulating a
non-conducting bulk.
%  The surface bands are readily observed in
%angle-resolved photoemission experiments, and several recent articles,
%including Refs.~\cite{kn:hasan2010,kn:moore2010,kn:qi2011} review the
%early theoretical and experimental literature.  
%Our primary interest is in the electronic properties of
%these surface states as would be seen in a typical transport or
%scanning probe measurement.  
From this perspective, a major
shortcoming of the early experiments was that carrier doping levels
were sufficiently high, resulting in the bulk bands being metallic, as
opposed to insulating.  This made it almost impossible to definitively
separate the properties of the surface state from that of the bulk,
and the system did not behave as a TI because the bulk is conducting
rather than insulating. This leads to the problem that while
spectroscopic measurements such as ARPES demonstrate the clear
existence of the expected surface topological bands, transport
measurements have been difficult to interpret due to the coexistence
of both bulk and surface conduction.  

%We are therefore encouraged by two very recent experimental studies, 
%one using scanning tunneling microscopy (STM) and the 
%second doing transport on
%Bi$_2$Se$_3$~\cite{kn:beidenkopf2011,kn:kim2011}.  
It is, therefore, encouraging that several recent experimental studies 
\cite{kn:kim2011,kn:hong2011,kn:kong2011,kn:steinberg2011} report the
direct observation of 2D surface states in transport measurement.
These reports claim to observe 
the electronic properties of the surface state with energy close to the 
topologically protected 
band crossing point (also called the Dirac point).  Our previous work on 
the electronic properties of graphene~\cite{kn:dassarma2011a} would 
lead us to expect that 
close to the Dirac point the energy landscape and the spatial
electronic structure would become highly inhomogeneous, 
breaking the surface into puddles of electrons and holes.  The charge
inhomogeneity  
would also result in a low density plateau in the conductivity with a
non-vanishing minimum  
conductivity at the Dirac
point~\cite{kn:dassarma2011a,kn:hwang2006b,kn:adam2007a,kn:rossi2008b}. 
Given the highly disordered nature of the experimental systems, we
expect the electron-hole puddles to completely dominate the 2D
transport and electronic structure near the Dirac point.

This physically intuitive picture of inhomogeneous electron and hole
puddles has been highly successful in understanding both transport
experiments~\cite{kn:tan2007,kn:chen2008} and scanning probe
experiments~\cite{kn:adam2011} in graphene.  The goal of this work is
to understand the role of a locally fluctuating carrier density on the
electronic screening and conductivity of the $2D$ surface states in
these new $3D$ TI materials.  Our main finding 
is that band asymmetry, which can almost
always be ignored when studying graphene, plays a crucial role in
TI experiments.  One striking example
is that in some cases, the TI surface conductivity
does not even have a minimum conductivity in the vicinity of the
band-crossing point, making the physics very different from that of
graphene.
Comparing our theory with the existing Bi$_2$Se$_3$ experiments, 
\cite{kn:kim2011,kn:hong2011,kn:kong2011,kn:steinberg2011}
we also conclude that the current systems have very large background
disorder making it difficult to access the physics of the Dirac point.

%%%%%%%%%%%%%%%%%%%%%%%%%%%%%%%%%%%%%%%%%%%%%%%%%%%%%%%%%
%%%%%%%%%% FIG. 1
\begin{figure*}[ht!]
\begin{center}$
\begin{array}{cc}
\includegraphics[width=3.2in]{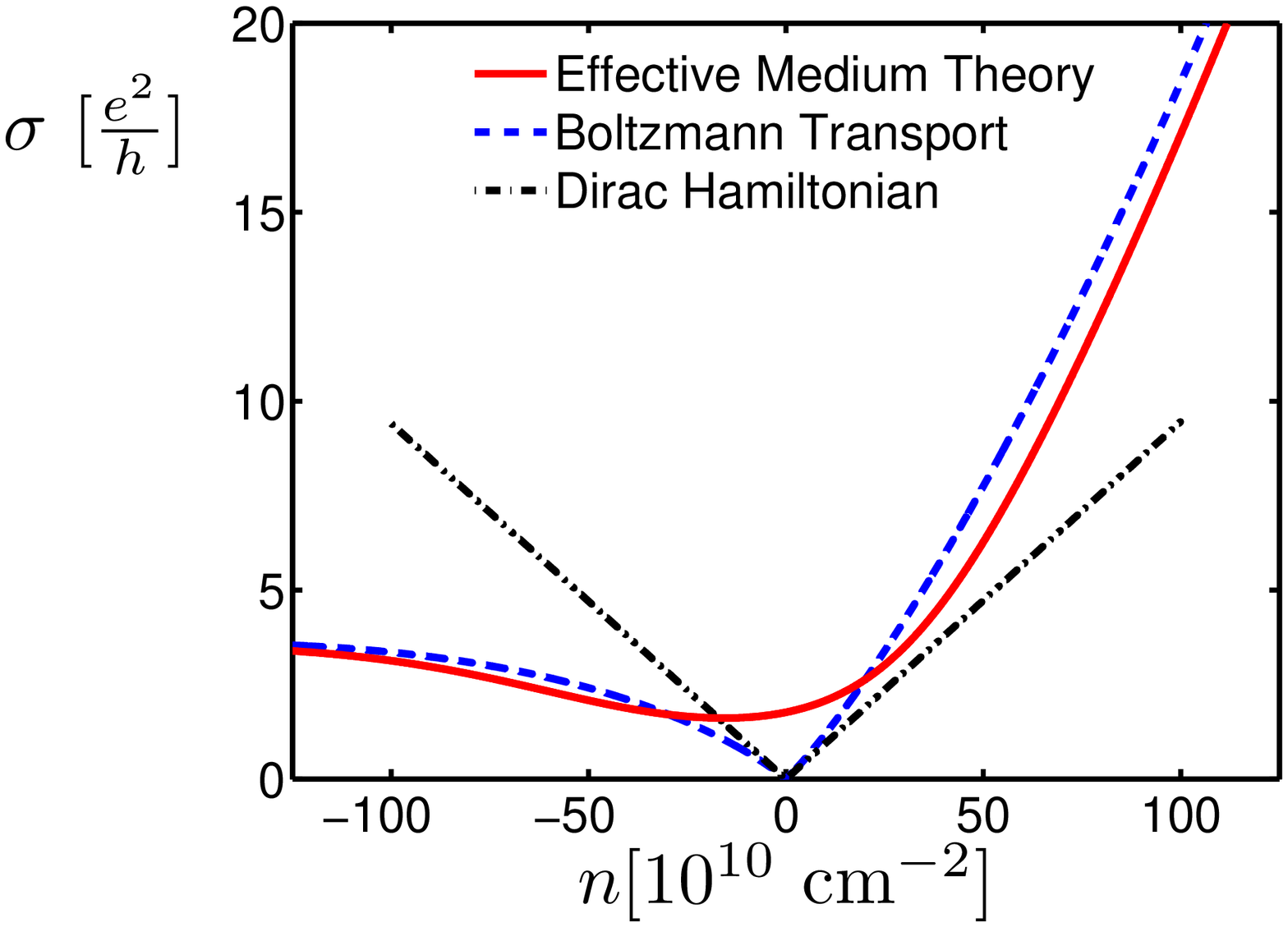} &
\includegraphics[width=3.2in]{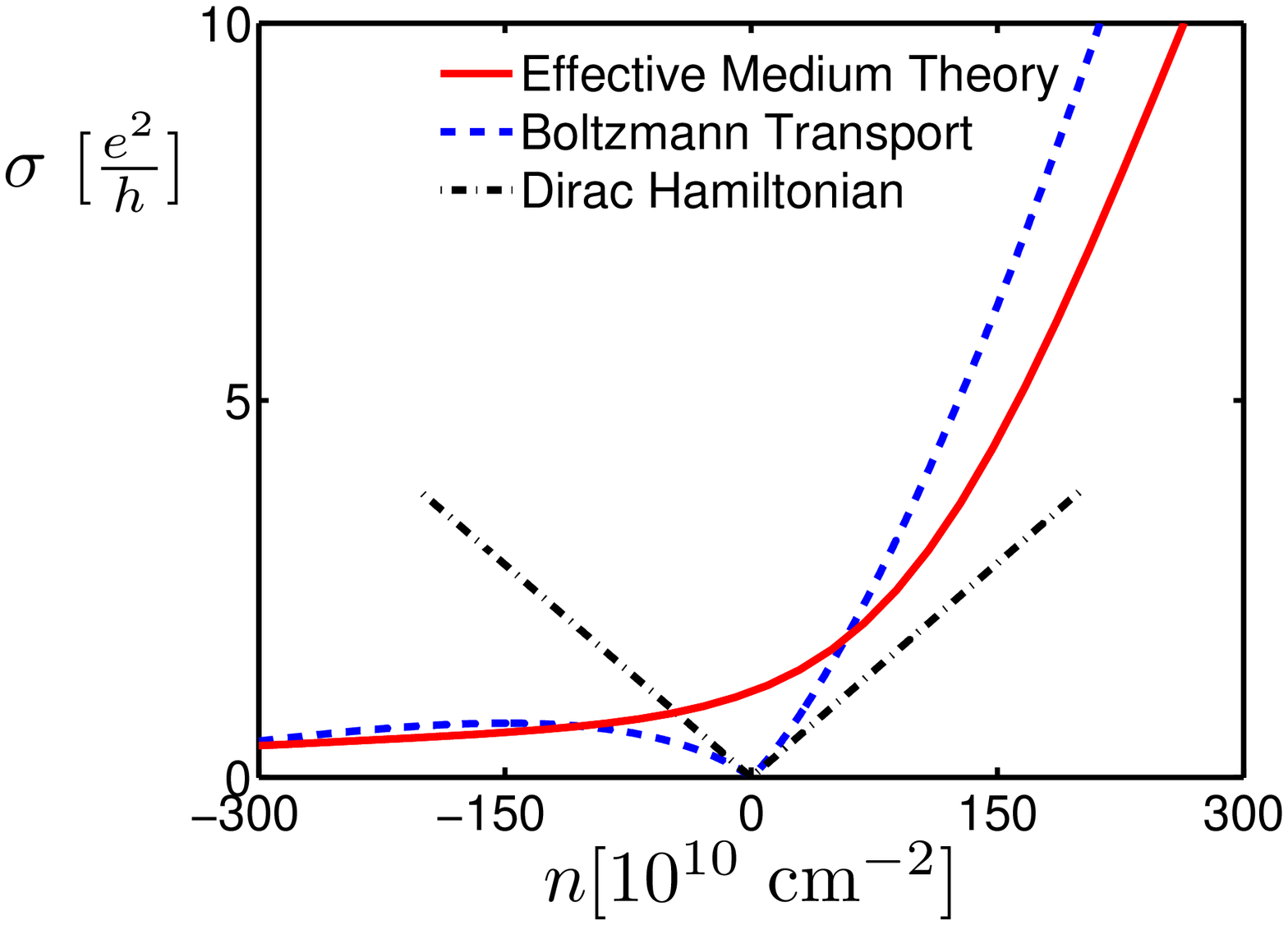}
\end{array}$
\end{center}
\caption{\label{Fig:Conductivity} 
(Color online) Theoretical calculations for dependence of conductivity
  on carrier density using parameters $r_s = 0.1$ and $n_0 = 2 \times
  10^{13}~{\rm cm}^{-2}$.  The left panel has impurity concentration
  $n_{\rm imp} = 10^{13}~{\rm cm}^{-2}$, and the right panel is dirtier with $n_{\rm imp} = 5 \times 10^{13}~{\rm
    cm}^{-2}$. Dashed blue lines show the Boltzmann transport result
  described in Eq.~(\ref{Eq:Boltzmann}), and dash-dotted black lines are
  for Dirac bands i.e. $m^* \rightarrow \infty$ in
  Eq.~(\ref{Eq:Hamiltonian}).  The red solid line is the result of the
  effective medium theory calculation obtained by solving
  Eq.~(\ref{Eq:EMT}) using $d=0.1~{\rm nm}$.  The left panel has 
a conductivity minimum in the vicinity of the Dirac point, while for the 
right panel the conductivity monotonically increases as the carriers are 
tuned from holes (negative $n$) to electrons (positive $n$).}
\end{figure*}

The starting point for our calculation is to make some reasonable approximation 
for the band structure of the topological surface state.  One approach
\cite{kn:zhang2009c,kn:liu2010} would be to perform {\it ab initio}
calculations and fit that data to the most 
general model Hamiltonian allowed by symmetry.  However, we find significant discrepancy between 
the electronic structure calculations and the photoemmision (ARPES) experiments.  As a result, a reasonable comparison with experiment would require us fine-tuning a model Hamiltonian with 12 
parameters.  While this procedure could be done, it would be unnecessarily cumbersome and obscure any 
physical insight.  Instead, we follow the minimal model proposed
earlier~\cite{kn:culcer2010} in the literature mimicking the full {\it
  ab initio} band structure   
\begin{equation}
{\cal H}({\bf k}) = \frac{\hbar^2 k^2}{2 m^*} + \hbar v_{\rm F} \left( k_x \sigma_y - k_y \sigma_x \right),
\label{Eq:Hamiltonian}
\end{equation}
where $(\sigma_x, \sigma_y)$ is a $2D$ vector of Pauli matrices,
${\bf k} = (k_x,k_y)$ is the 2D wave vector,
$v_{\rm F}$ is the Fermi velocity of the Dirac bands, and effective
mass $m^*$ characterizes the degree of asymmetry between the electron
and hole bands.  Estimates for the values of these two parameters
for Bi$_2$Se$_3$ vary widely in the experimental literature.  For
example, values for $v_{\rm F}$ vary from $2 \times
10^{5}~{\rm m/s}$~\cite{kn:beidenkopf2011} to $6.4 \times 10^{5}~{\rm m/s}$~\cite{kn:kim2011}, and
measured values for $m^*$ vary from $0.11~{\rm m}_{\rm e}$~\cite{kn:analytis2010}
(${\rm m}_{\rm
  e}$ is the electron mass) to $0.32~{\rm m}_{\rm e}$~\cite{kn:taskin2011}.  This
situation should be contrasted with graphene, where $v_{\rm F}$ is the
single band parameter, and most experimental reports agree on its
value to within $5$ percent~\cite{kn:dassarma2011a}.
We use the two parameter ($m^*$ and $v_F$) model of
Eq.~(\ref{Eq:Hamiltonian}) in the current work.

Since we are concerned with the screening properties of electrons, it
is useful to define an interaction parameter $r_s = e^2/(\kappa \hbar
v_{\rm F})$, where we reiterate that throughout this work, $v_{\rm F}$
is the parameter in Eq.~(\ref{Eq:Hamiltonian}) characterizing the
Dirac-like bands; only at low carrier density does it coincide with
the Fermi velocity.  Here $\kappa$ is approximately half the
dielectric constant of the bulk Bi$_2$Se$_3$ insulator (whose reported
value varies from around $30$~\cite{kn:beidenkopf2011} to about
$55$~\cite{kn:butch2010}).  We certainly expect that as more
experiments on TI materials become available, these
parameters will become better known both for Bi$_2$Se$_3$ and other
related materials.  To make our theory more compact, we formulate
everything in terms of $r_s$ and a characteristic density $n_0 = (m^*
v_{\rm F})^2/(4 \pi \hbar^2)$, where reasonable values of $r_s$ are in the
range $0.05$ to $0.5$, and reasonable values of $n_0$ are from
$10^{11}~{\rm cm}^{-2}$ to $3 \times 10^{13}~{\rm cm}^{-2}$.
We note that $n_0$ is an important parameter characterizing the
deviation of the system (for $n >n_0$) from purely Dirac-like behavior
--- in graphene, $n_0$ is very large.

The Thomas-Fermi screening theory for electrons specifies that all
external potentials  
are screened by a surface 2D dielectric function $\epsilon(q) = 1 +
q_{\rm TF}/q$.  For  
the Hamiltonian in Eq.~(\ref{Eq:Hamiltonian}), we find 
\begin{equation}
q_{\rm TF} = \frac{r_s k_{\rm F}}{1 + {\rm sgn}(n) \sqrt{|n|/n_0}} = \eta r_s k_{\rm F},
\end{equation}
where ${\rm sgn}$ is the signum function, and we use the convention that electrons 
have  ${\rm sgn}(n) =+1$ while holes have ${\rm sgn}(n) =-1$.  For both electrons 
and holes, we have $k_{\rm F} = \sqrt{|4 \pi n|}$.  Note that $q_{\rm TF}$ diverges 
for $n \rightarrow -n_0$, implying perfect screening associated with the diverging 
density of states in Eq.~(\ref{Eq:Hamiltonian}) arising from the
quadratic dispersion and the band asymmetry. It turns out that the
theory can be completely characterized by the two parameters $r_s$ and
$n_0$, rather than the three microscopic parameters $m^*$, $v_F$, and
$\kappa$.   

Our numerical analysis using the full random phase approximation (RPA) 
%with a slightly modified Hamiltonian (discussed below) 
shows that the
Thomas-Fermi analysis is accurate provided we restrict the carrier
density for holes to $|n| < |n_0|$.  No such restriction is required
for electrons.  Within the Boltzmann transport approximation, the
conductivity $\sigma = (k_{\rm F} \ell)(e^2/2 h)$, where $\ell =
v_{\rm F} \tau/\eta$ is the mean-free path. The scattering time is
calculated within Born approximation as
\begin{equation}
\frac{\hbar}{\tau} = \pi n_{\rm imp} \sum_{\bf k'} \left| \frac{v(q)}{\epsilon(q)} \right|^2 \sin^2[\theta_{kk'}] \delta(\varepsilon_k - \varepsilon_{k'}),
\end{equation}
where $n_{\rm imp}$ is the surface density of random charged
impurities, $\varepsilon_k$ is the carrier energy, $\theta_{kk'}$ is
the scattering angle between wave vectors ${\bf k}$ and ${\bf k'}$,
and $v(q)/\epsilon(q)$ is the Fourier  
transform of the screened 
impurity potential.  For the purpose of this calculation we assume that the 
dominant scatterers are long-ranged Coulomb impurities 
(with an average 2D density of $n_{\rm imp}$ placed an average distance of $d$
away from the TI surface) 
although our formalism can easily 
be generalized to other types of impurities.  For these charged
impurities, the conductivity can be calculated analytically, giving 
\begin{eqnarray}
\label{Eq:Boltzmann}
\sigma_B[n, n_0, r_s] &=& \frac{1}{8} \frac{e^2}{h} \frac{n}{n_{\rm imp}} \frac{1}{F_1[\eta r_s/2]}, \\
\frac{F_1[x]}{x^2} &=& \frac{\pi}{4} + 3x -\frac{3x^2 \pi}{2} + x (3x^2 -2) \frac{\arccos[1/x]}{\sqrt{x^2 -1}}, 
          \nonumber \\
\eta[n/n_0] &=& \frac{\sqrt{n_0}}{\sqrt{n_0} + {\rm sgn}(n) \sqrt{|n|}}. \nonumber 
\end{eqnarray}
The results for the Boltzmann transport theory are shown in
Fig.~\ref{Fig:Conductivity}.  One immediately observes that the
asymmetry in the conductivity is quite pronounced compared to a linear
Dirac dispersion ($m^* = \infty$; also shown).  The electron branch
has a much larger conductivity with $\sigma_B(n)$
being super-linear, while the hole branch has a much lower sub-linear
$\sigma_B(n)$.
% (and for realistic parameters remains on order of the
%conductance quantum). 
This pronounced asymmetry between electron and hole transport,
following directly from the band asymmetry of
Eq.~(\ref{Eq:Hamiltonian}), is a characteristic feature of 2D TI
transport.

At low carrier density, the disorder induced fluctuations in carrier
density become larger than the average carrier density.  In
particular, when the average carrier density vanishes with the
chemical potential at the Dirac point, one might
expect that the electronic properties of the system are determined by
the typical carrier density inside the electron and hole puddles.  For
example, one could define a carrier density distribution function
$P[n]$, and the condition of zero average carrier density implies the
vanishing of the first moment of $P[n]$.  The second moment of the
carrier density distribution $n_{\rm rms}$ would then determine how
the carriers in this inhomogeneous system screen any external
potential, including the impurity potential that induced the fluctuations to
begin with.  This implies that the density fluctuations need to be
calculated self-consistently\cite{kn:adam2007a}. 
We calculate the properties of this inhomogeneous system by assuming a global screening function that
depends on the impurity profile only through an effective carrier
density $n_{\rm eff}$ (where knowing $n_{\rm eff}$, one can then
calculate all moments of $P[n]$ including $n_{\rm rms}$, e.g. for
Dirac fermions, $n_{\rm rms} \approx \sqrt{3} n_{\rm eff}$). 

%%%%%%%%%%%%%%%%%%%%%%%%%%%%%%%%%%%%%
%%%%%%  FIG. 2
\begin{figure}
\includegraphics[width=3.2in]{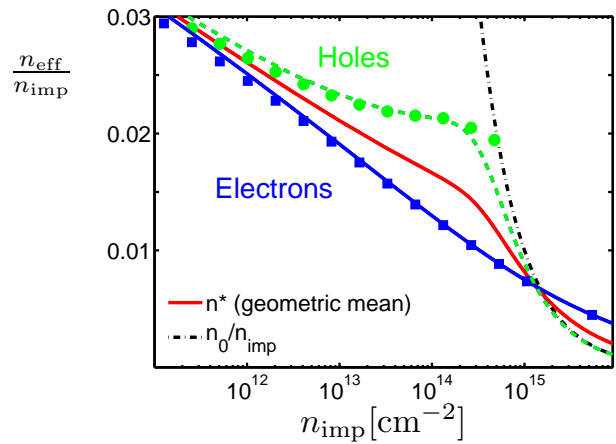}
\caption{\label{Fig:Neff} 
(Color online) Theoretical calculations for disorder 
induced carrier density $n_{\rm eff}$ as a function of 
impurity concentration $n_{\rm imp}$ using parameters $r_s = 0.1$, $n_0 = 10^{13}~{\rm cm}^{-2}$, 
and $d= 0.1~{\rm nm}$.  The curves were calculated using the Thomas-Fermi screening 
theory, while the circles (holes) and squares (electrons) show the RPA.
For 
electrons, the two approximations agree to within our numerical accuracy.  For holes, they agree 
only for $n_{\rm eff} < n_0$ (shown as a dotted line).  One implication of this result is that charge neutrality point will be 
distinct from the Dirac point.}
\end{figure}

This effective carrier density $n_{\rm eff}$ is nothing other than a
measure of the typical carrier density inside the electron and hole
puddles.  After obtaining $n_{\rm eff}$, we can then compute other properties of the Dirac point 
including its conductivity (that can be measured in a transport experiment) or its
density-density correlation function (measured in STM).  Calculating $n_{\rm eff}$ is
therefore a central result of this work.  We do this by requiring that the density induced by the
second moment of the screened disorder potential is precisely the same
as the density entering the global screening function.  Applying this 
procedure to Eq.~(\ref{Eq:Hamiltonian}), and defining the 
dimensionless variable $y = |n_{\rm eff}|/n_0$, we derive a system 
of equations that can be easily solved numerically
\begin{eqnarray}
\label{Eq:SCA}
\frac{y^2}{4} + y + s y^{3/2} &=& A C_0\left[ \frac{ B \sqrt{y}}{1 + s \sqrt{y}} \right], \\
  C_0[x] &=& \partial_x \left[ x e^{x} \int_x^\infty t^{-1} e^{-t} dt \right], \nonumber \\
A = \frac{1}{2} \frac{n_{\rm imp}}{n_0} r_s^2, && B = 2 r_s d \sqrt{4 \pi n_0}. \nonumber
\end{eqnarray}
\noindent 
Here, $s= {\rm sgn}(n)$ denotes the electron ($s=+1$) and hole
($s=-1$) bands.  
%Since the bare potential of a 
%charged impurity diverges when the
%impurity is at the surface, we need to make the physical assumption
%that the impurities are 
%on average displaced by a distance $d$ from the surface of the TI.  

In Fig.~\ref{Fig:Neff} we show the effective carrier density obtained
using this self-consistent method.  Notice that the induced carrier density is
different for electrons and for holes except in the limiting case
$n_{\rm imp} \ll n_0$ [where Eq.~(\ref{Eq:Hamiltonian}) gives a
symmetric linear Hamiltonian].  This asymmetry implies that the charge
neutrality point (where number of electrons and holes are equal) does
not necessarily coincide with the Dirac point (the crossing point
between electron 
and hole bands).  The numerical value of $n_{\rm eff}/n_{\rm imp}$ is
rather small (less than 0.03).  This not only guarantees convergence
of the theory, but also has important implications for experiments.
Since $n_0$ and $n_{\rm imp}$ are comparable in current experiments,
if $n_{\rm eff}/n_{\rm imp}$ is not small, this would imply that the
energy scale associated with the disorder-induced inhomogeneity would
be comparable to the bulk band-gap, and there would be no hope of
observing any physics associated with the topological surface state.
We note that in graphene $n_{\rm eff} \sim n_{\rm imp}$ in contrast to 2D
TI transport where we are finding $n_{\rm eff} \ll n_{\rm imp}$.

%\begin{figure*}[ht!]
%\begin{center}$
%\begin{array}{cc}
%\includegraphics[width=2.5in]{Figure_kappa_1.eps}  & \includegraphics[width=2.3in]{Figure_nimp_1.eps} \\
%\includegraphics[width=2.5in]{Figure_vf_1.eps}  & \includegraphics[width=2.5in]{Figure_d_1.eps}
%\end{array}$
%\end{center}
%\caption{\label{Fig:panel} 
%(Color online) Minimum conductivity as a function of system parameters (a) impurity density, (b) Fermi velocity%, 
%(c) dielectric constant, and (d) impurity distance.  The values for the parameters not being varied are: 
%$n_{\rm imp} = 5\times 10^{13}~{\rm cm}^{-2}, v_{\rm F} = 4.5 \times 10^{5}~{\rm m/s}, \kappa = 50, d = 0.1~{\r%m nm}$,
%and $m^* = 0.2$.  A zero value for $\sigma_{\rm min}$ means that there was no local minimum in the conductivity% 
%in the vicinity of the Dirac point.  In these cases (in contrast to Fig.~\protect{\ref{Fig:Conductivity}}) the
%disordered averaged EMT conductivity increases monotonically as one increases the density, and the minimum 
%conductivity values occurs at the lowest attainable gate-voltage corresponding to the highest hole density.}
%\end{figure*}

\begin{figure*}[ht!]
\begin{center}$
\begin{array}{cccc}
\includegraphics[width=1.6in]{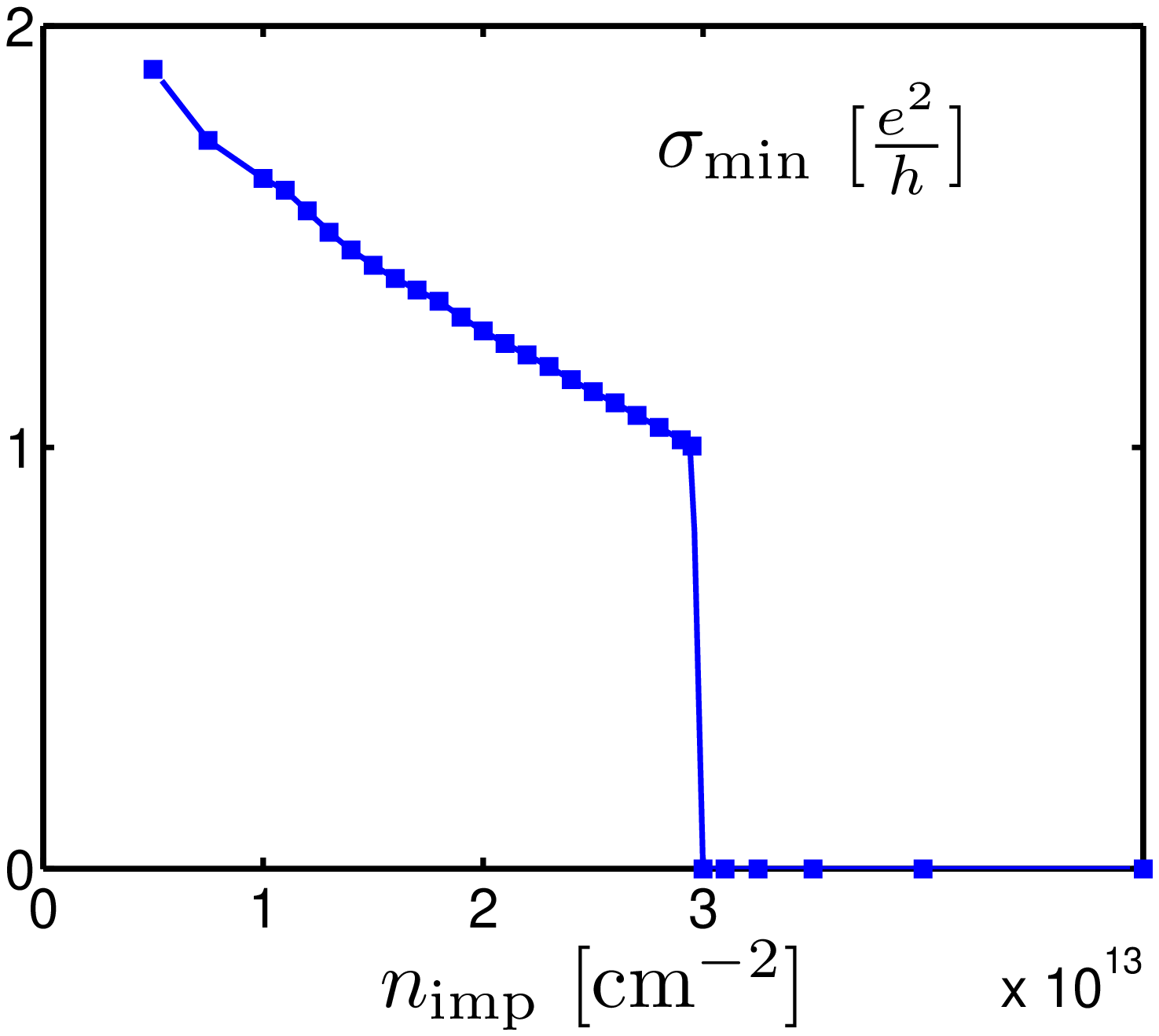}  &
\includegraphics[width=1.65in]{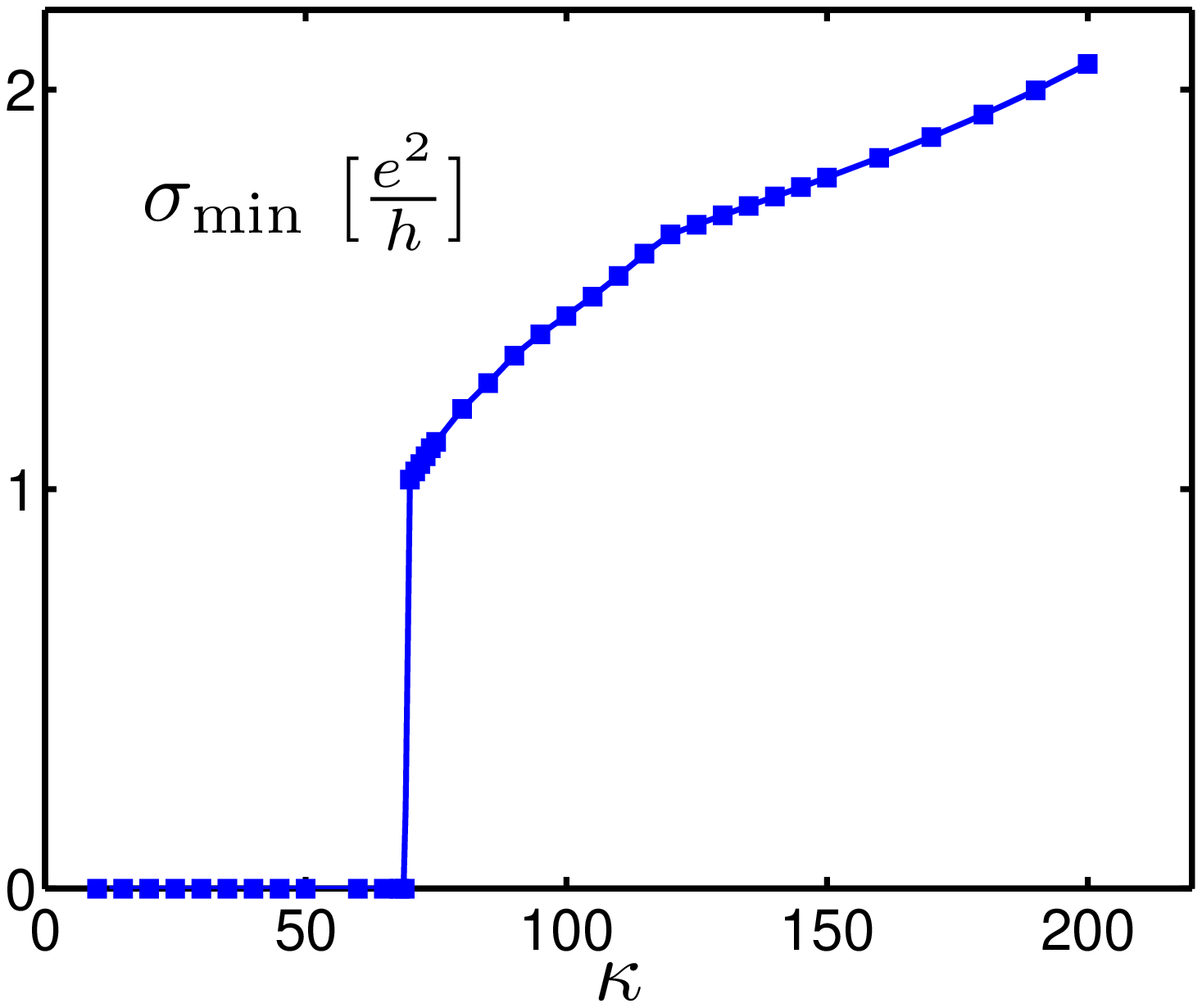}  & 
\includegraphics[width=1.6in]{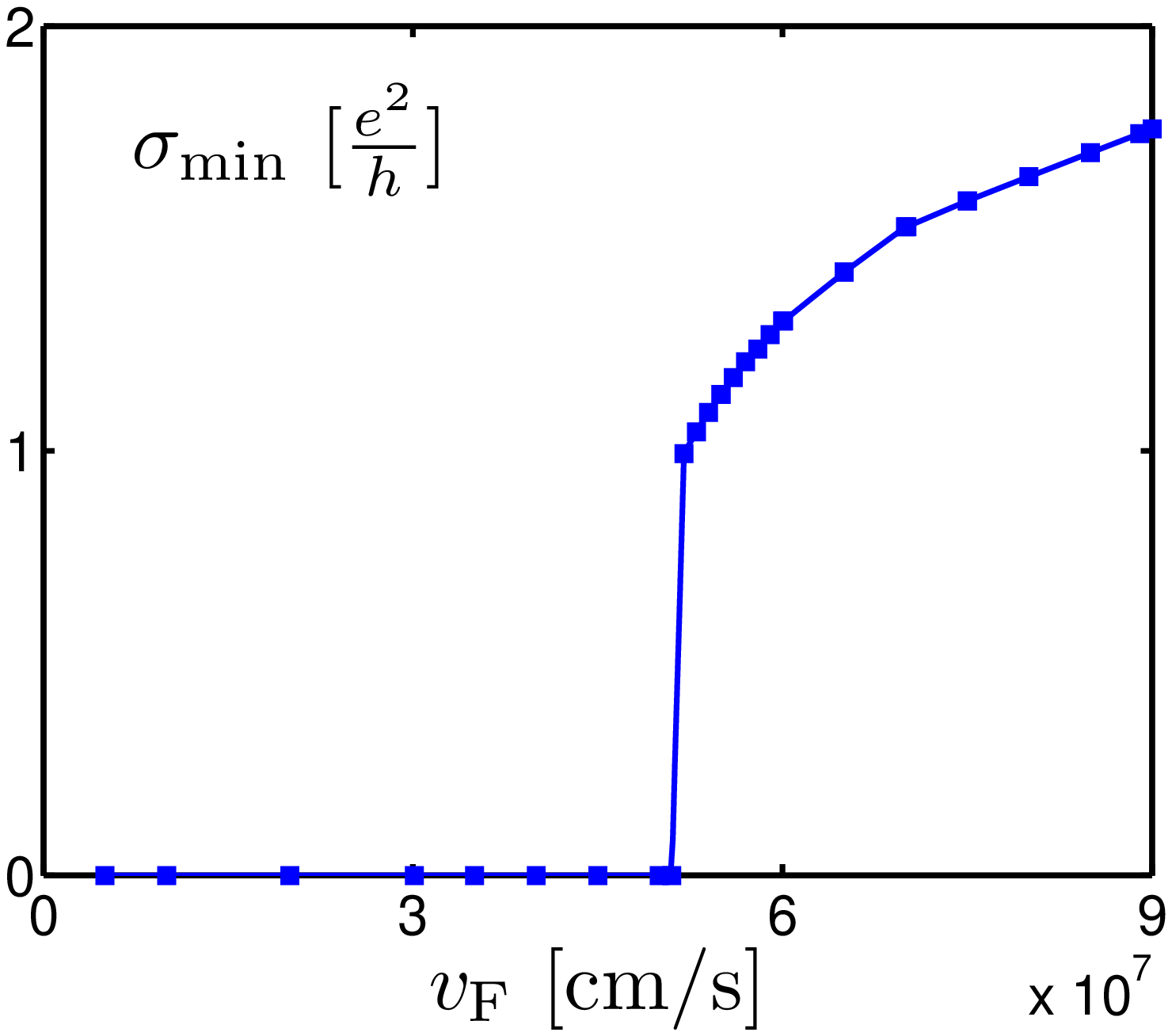}  & 
\includegraphics[width=1.65in]{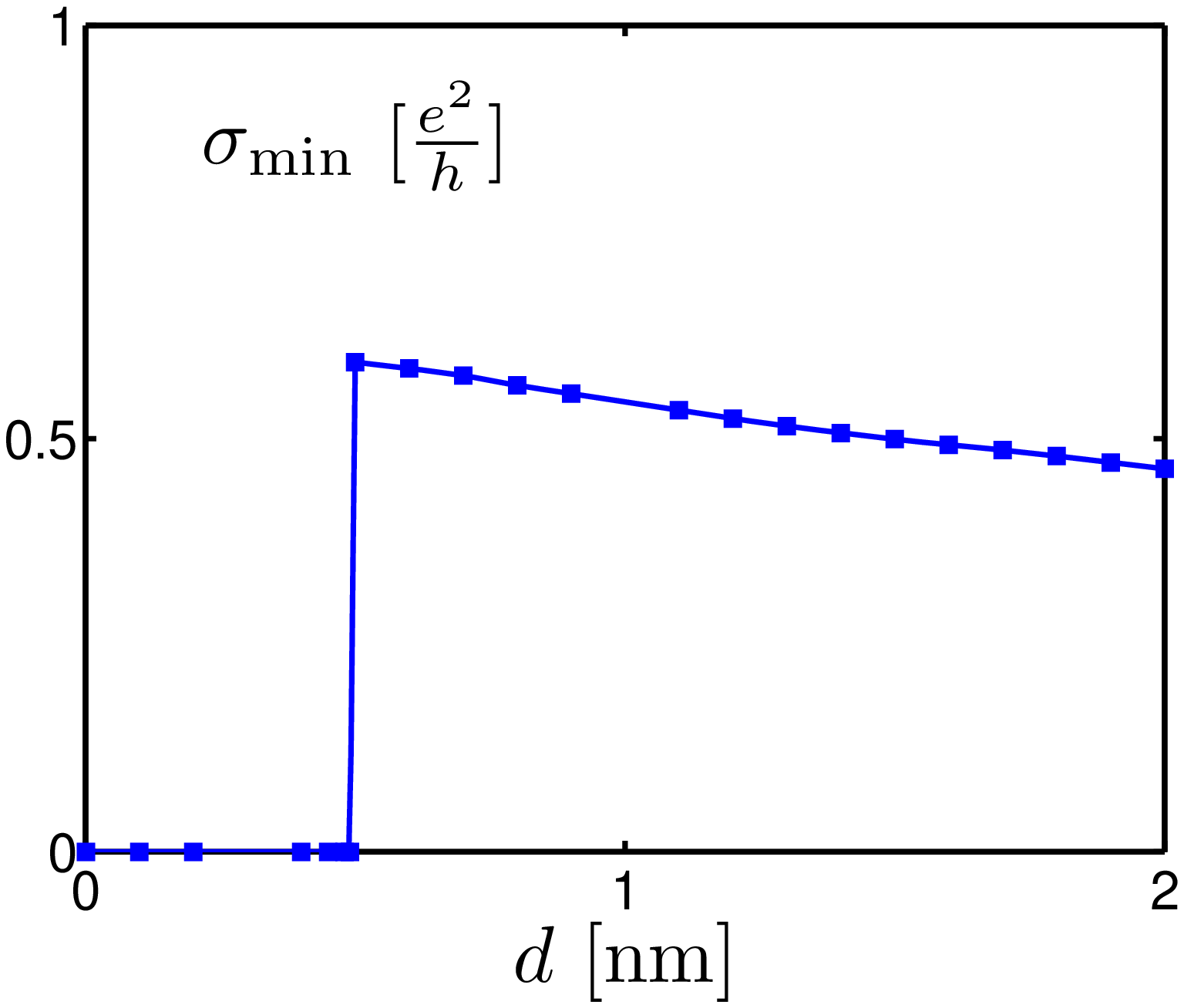}
\end{array}$
\end{center}
\caption{\label{Fig:panel} 
(Color online) Minimum conductivity as a function of system parameters
  (a) impurity concentration $n_{\rm imp}$, (b) dielectric constant $\kappa$, (c) low energy Fermi velocity
  $v_{\rm F}$, and (d) impurity distance $d$.  The values for the
  parameters not being varied are: $n_{\rm imp} = 5\times 10^{13}~{\rm
    cm}^{-2}, v_{\rm F} = 4.5 \times 10^{5}~{\rm m/s}, \kappa = 50, d
  = 0.1~{\rm nm}$, and $m^* = 0.2$.  A zero value for $\sigma_{\rm
    min}$ means that there was no local minimum in the conductivity in
  the vicinity of the Dirac point. Solid curves are a guide to the eye.}
\end{figure*}

We now proceed to calculate the electronic transport in this
inhomogeneous puddle-dominated carrier density landscape.
The procedure is to first assume that the local conductivity can be
calculated from the local carrier density  
using Eq.~(\ref{Eq:Boltzmann}).  Then, an effective medium theory (EMT)
can be used to calculate the global conductivity from this
distribution of spatially fluctuating local conductivities \cite{kn:rossi2008b} 
\begin{equation}
\int dn P[n] \frac{\sigma_B(n) - \sigma_{\rm EMT}}{\sigma_B(n) + \sigma_{\rm EMT}} = 0.
\label{Eq:EMT}
\end{equation}
\noindent The EMT assumes that the dominant
contribution to the resistivity arises from scattering inside the
electron and hole puddles, and not across the puddles.  This is a
reasonable assumption because the cross puddle backscattering is
suppressed due to spin conservation.  To simplify the theory we assume
that $P[n]$ is Gaussian, with variance $\sqrt{3} n^*$, where $n^* =
\sqrt{n_{\rm eff}^e n_{\rm eff}^h}$ is the geometric mean of the
electron and hole density fluctuations (see Fig.~\ref{Fig:Neff}).  Our
results are shown in Fig.~\ref{Fig:Conductivity}.  An important result
of performing the EMT average using an asymmetric conductivity is that
the carrier density at which the conductivity is minimum is distinct
from both the charge neutrality point and Dirac point, further
complicating the analysis and interpretation of experiments.  This is
illustrated dramatically in the right-hand panel, where there is no
conductivity minimum in the vicinity of the Dirac point.  Rather, the
conductivity increases monotonically from $\sigma = 0$, for $n
= -n_0$ to $\sigma_B(|n|)$ for $n \gtrsim n_0$.  Indeed, in the
experiments of Ref.~\cite{kn:kim2011}, the most disordered sample 
does not show a (clear) minimum conductivity, while other devices do have a 
local $\sigma_{\rm min}$ close to the Dirac point, similar to that
shown in the left-hand panel of Fig.~\ref{Fig:Conductivity}.
The possible non-existence of a conductivity minimum in 2D TI
transport is a new qualitative prediction of our theory.

We have explored how $\sigma_{\rm min}$ depends on a variety of physical 
parameters.  In most cases, we find that the minimum conductivity {\it
  increases}  
with {\it increasing} $\kappa$, $v_{\rm F}$, $m^*$ and sample purity ($n_{\rm imp}^{-1}$).
However, there is also a large range of parameter space where there is no conductivity 
minimum associated with the Dirac point (we denote this case as $\sigma_{\rm min} = 0$). 
We illustrate our results in Fig.~\ref{Fig:panel}, where we show the dependence of 
$\sigma_{\rm min}$ on $\kappa$, $v_{\rm F}$, and $d$.  As seen in the figure, in each of these cases, 
the crossover between the presence or absence of a well defined $\sigma_{\rm min}$ occurs quite sharply
as a function of the parameter being varied.

The analytic results provided in Eq.~(\ref{Eq:SCA}) were done using the Thomas-Fermi (TF) approximation.
We have also done calculations using the full static RPA.
We find in general, as shown in Fig. 2, very good agreement between
the TF and RPA results with the RPA results being obtained completely
numerically.

\begin{figure}
\includegraphics[width=2.5in]{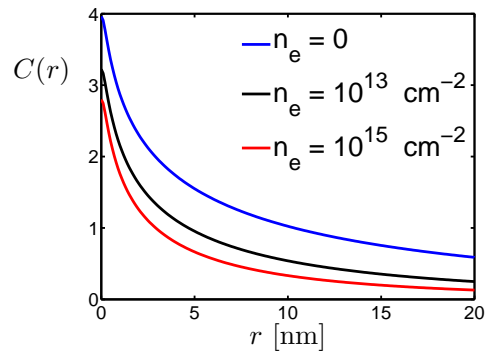}
\caption{\label{Fig:Xi} 
(Color online) Theoretical calculations for potential correlation fluctuation 
$C(r) = \langle (V(r) - {\bar V})(V(0) - {\bar V}) \rangle$ for the electron band using the 
same parameters as in Fig.~\protect{\ref{Fig:Conductivity}}b.  The y-axis is normalized 
by $(r_s \hbar v_{\rm F} \sqrt{\pi n_{\rm imp}})^2$. From top to bottom, the curves represent average electron doping of
$ n =  0$, $10^{13}~{\rm cm}^{-2}$, and $10^{15}~{\rm cm}^{-2}$.}
\end{figure}

Finally, we turn our attention to the recent
experiments which have directly observed the electron-hole puddles on
the Bi$_2$Se$_3$ surface \cite{kn:beidenkopf2011}.  In addition to being a local
probe, by tuning the bias voltage between the tip and the sample,
these experiments can probe features in the density of states away
from the Fermi energy.  In current TI materials,
this is especially important because unintentional doping pushes the
Fermi energy far from the Dirac point.  For each position in space,
one can map out the energy of the Dirac point $E_D$.  Shifts of $E_D$ 
away from the average potential ${\bar V}$ gives the screened disorder potential landscape whose 
autocorrelation function is $C(r) = \langle (V(r) - {\bar V})(V(0) - {\bar V}) \rangle$.
Theoretical calculations for the correlation function are shown in Fig.~\ref{Fig:Xi} for 
different (average) electron carrier densities $n$ using the Thomas-Fermi approximation 
and solving for the total density $n_{\rm tot} = n + n_{\rm eff}$ self-consistently.  The effect of electron 
doping is to slightly reduce both the magnitude of the potential fluctuations $C(0)$ as well as the spatial 
correlation length of the puddles.  

Our predictions for screening and carrier transport in disordered
2D TI surface states should be 
testable in future experiments.  In particular, the band asymmetry
gives rise to interesting and qualitatively novel
effects, e.g., causing the Dirac point, the charge neutrality point,
and the minimum conductivity to occur at different carrier densities.  
%Such properties makes the theoretical study 
%of topological insulators far richer than other chiral electron systems like graphene.
In addition, electron and hole 2D transport in TIs should manifest
strong asymmetry, and in some situations with strong disorder
(i.e., large $n_{imp}$ or small $d$) there may not be any minimum
conductivity plateau associated with the Dirac point.

This work is supported by US-ONR and NRI-SWAN.  
% \vspace*{-0.25cm}
% \bibliography{/home/electron_physics/shaffiq/Files/LaTex/shaffiquebib.bib}

\end{document}